\documentclass[prb,aps,a4paper]{revtex4}
\usepackage[latin1]{inputenc}
\usepackage{amsmath}
\usepackage{amssymb}
\usepackage{amscd}
\usepackage{graphicx}
\usepackage{subfigure}

\def\sump{\sideset{}{'}\sum}

\def\O{{\mathcal O}}

\def\Z{\mathbb{Z}}

\def\mpt{\,\text{.}}
\def\mcm{\,\text{,}}

\newlength\picturewidth
\newlength\pictureheight
\newcommand\algo[1]{{\bf #1}}
\begin{document}

\title{Electrostatics in Periodic Slab Geometries I}
 
\author{Axel Arnold}
\email{arnolda@mpip-mainz.mpg.de}
\author{Jason de Joannis}
\email{joannis@mpip-mainz.mpg.de}
\author{Christian Holm}
\email{holm@mpip-mainz.mpg.de}
\affiliation{Max-Planck-Institut f\"{u}r Polymerforschung,
Ackermannweg 10, 55128 Mainz,
Germany}                                                                       \date{\today}



\begin{abstract}
  We propose a new method to sum up electrostatic interactions in 2D
  slab geometries. It consists of a combination of two recently
  proposed methods, the 3D Ewald variant of Yeh and Berkowitz, J.
  Chem. Phys. 111 (1999) 3155, and the purely 2D method MMM2D by
  Arnold and Holm, to appear in Chem. Phys. Lett. 2002. The basic idea
  involves two steps. First we use a three dimensional summation
  method whose summation order is changed to sum up the interactions
  in a slab-wise fashion. Second we subtract the unwanted interactions
  with the replicated layers analytically. The resulting method has
  full control over the introduced errors. The time to evaluate the
  layer correction term scales linearly with the number of charges, so
  that the full method scales like an ordinary 3D Ewald method, with
  an almost linear scaling in a mesh based implementation. In this
  paper we will introduce the basic ideas, derive the layer correction
  term and numerically verify our analytical results.

\end{abstract}
\pacs{73, 41.20 Cv}
 \preprint{ELC I}
\maketitle

%
  


 
 


\section{Introduction}
The calculation of long range interactions due to Coulomb, gravitational, or
dipolar particles is of broad interest from the astrophysics, the biophysics
up to the solid state community.  These interactions present a formidable
challenge even to modern computers. Sophisticated methods such as fast
multipole methods, tree code algorithms, Poisson grid solvers, or Ewald mesh
methods, bring the complexity of N interacting particles down to an almost
linear scaling for three-dimensional periodic systems. Often, however, one is
interested in slab-like systems which are only periodic in two space
dimensions and finite in the third, for example in problems involving
electrolyte solutions between charged surfaces, proteins near charged
membranes, thin films of ferrofluids, Wigner crystals, charged films,
membranes, solid surfaces decorated with dipoles etc.

For such systems Ewald based formulas are only slowly convergent, have mostly
$\O(N^2)$ scalings and no ``a priori'' error estimates exist
\cite{widmann97a}.  Fast Ewald based methods have been recently put forward in
\cite{kawata01b}, and a non-Ewald method has been put forward in
Ref.\cite{lekner91a} that is based on a resummation of the force sum. However,
these methods are hampered due to non-controllable errors and an $\O(N^2)$
scaling respectively.  Recently we proposed a new method called
\algo{MMM2D}\cite{arnold01b,arnold01c} which has an $\O(N^{5/3})$ complexity
and full error control that is based on a convergence factor approach similar
to \algo{MMM}\cite{sperb98a}.  However, this will still only allow simulations
including up to a few thousand charges. There have been early attempts to use
a 3D Ewald sum for these slab problems. The main idea is to fill only parts of
the simulation box with charges and to leave some space empty, in an attempt
to decouple the interactions in the third dimension
\cite{shelley96a,spohr97a,yeh99a}.  Since each image layer is globally
neutral, one hopes that their interactions decay as they become more and more
distant, i.e. as the size of the gap is increased. In this way one could make
use of any advanced 3D Ewald implementation, see also Ref. \cite{minary02a}
for a variant of this idea.

In this paper we will follow the last suggestion and derive a term, called
electrostatic layer correction (\algo{ELC}), which subtracts the interactions
due to the unwanted layers.  The combination of that term with any three
dimensional summation method with slab--wise summation order will yield the
exact electrostatic energy.  Since the change in the summation order is done
by adding a very simple term, any three dimensional summation method with the
standard spherical summation order can be used.  The new term can be evaluated
easily in a time linear in the number of charges, hence the whole method
scales like the underlying standard summation method.  We develop also an
error formula for the maximal pairwise error in the energy and forces of the
layer correction term, hence the precision of this method can be tuned to any
desired value, when used in conjunction with other error estimates for the
standard summation method\cite{kolafa92a,deserno98b}.  In the first section we
will recapitulate the way how to correct the summation order via a modified
dipole term. In the second section we will derive the layer correction term,
and develop in the following section error estimates for its value. The
applicability of our method will be demonstrated by a numerical analysis in
the following section, and we end with our conclusion.

\section{Changing the summation order}
\begin{figure}[htbp]
  \begin{center}
    \includegraphics[width=5cm]{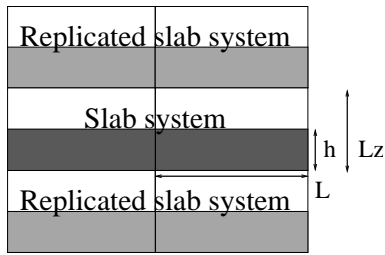}
    \caption{Schematic representation of a fully periodically
      replicated slab system}
    \label{fig:periodicity}
  \end{center}
\end{figure}

We consider a system of $N$ particles with charges $q_i$ and positions
$p_i=(x_i,y_i,z_i)$ that reside in a box of edges $L\times L\times h$, where
$h=\max_{i,j} |z_i - z_j|$ is the maximal $z$--distance of two particles.  The
basic idea is to expand this slab system in the non-periodic $z$--coordinate
to a system with periodicity in all three dimensions.  More precisely, the
original box of size $L\times L\times h$ is placed inside a box of size
$L\times L\times L_z$ where $L_z>>h$ sufficiently large. Then this box is
replicated periodically in all three dimensions. The result is a
three-dimensional periodic system with empty space regions (``gaps'') of
height $\delta:=L_z-h$ (see Fig.~\ref{fig:periodicity}).  $\delta$ will be
called gap size in the following.

Since the electrostatic potential is only finite if the total system is charge
neutral, the additional image layers (those layers above or below the original
slab system) are charge neutral, too.  Now let us consider the $n^{th}$ image
layer which has an offset of $nL_z$ to the original layer. If $nL_z$ is large
enough, each particle of charge $q_j$ at position $(x_j,y_j,z_j+nL_z)$ and its
replicas in the $x,y$-plane can be viewed as constituting a homogeneous
charged sheet of charge density $\sigma_j = \frac{q_j}{L^2}$.  The potential
of such a charged sheet at distance $z$ is $2\pi \sigma_j |z|$. Now we
consider the contribution from a pair of image layers located at $\pm nL_z$,
$n>0$ to the energy of a charge $q_i$ at position $(x_i,y_i,z_i)$ in the
central layer.  Since $|z_j - z_i| < nL_z$, we have $|z_j - z_i + nL_z| = nL_z
+ z_j - z_i$ and $|z_j - z_i - nL_z|= nL_z - z_j + z_i$, and hence the
interaction energy from those two image layers with the charge $q_i$ vanishes
by charge neutrality:
\begin{equation}
  \label{eq:parpltappr}
  2\pi q_i \sum_{j=1}^N \sigma_j(|z_j - z_i + nL_z| + |z_j - z_i - nL_z|)
  = 4\pi q_i nL_z \sum_{j=1}^N \sigma_j = 0 \mpt
\end{equation}
The only errors occurring are those coming from the approximation of
assuming homogeneously charged, infinite sheets instead of discrete
charges.  This assumption should become better when increasing the
distance $nL_z$ from the central layer.

\begin{figure}[htbp]
  \subfigure[Schematic view of the spherical summation order. $S$ is
  the length of the box offset.]{%
    \makebox[0.48\textwidth]{\includegraphics[width=0.3\textwidth]{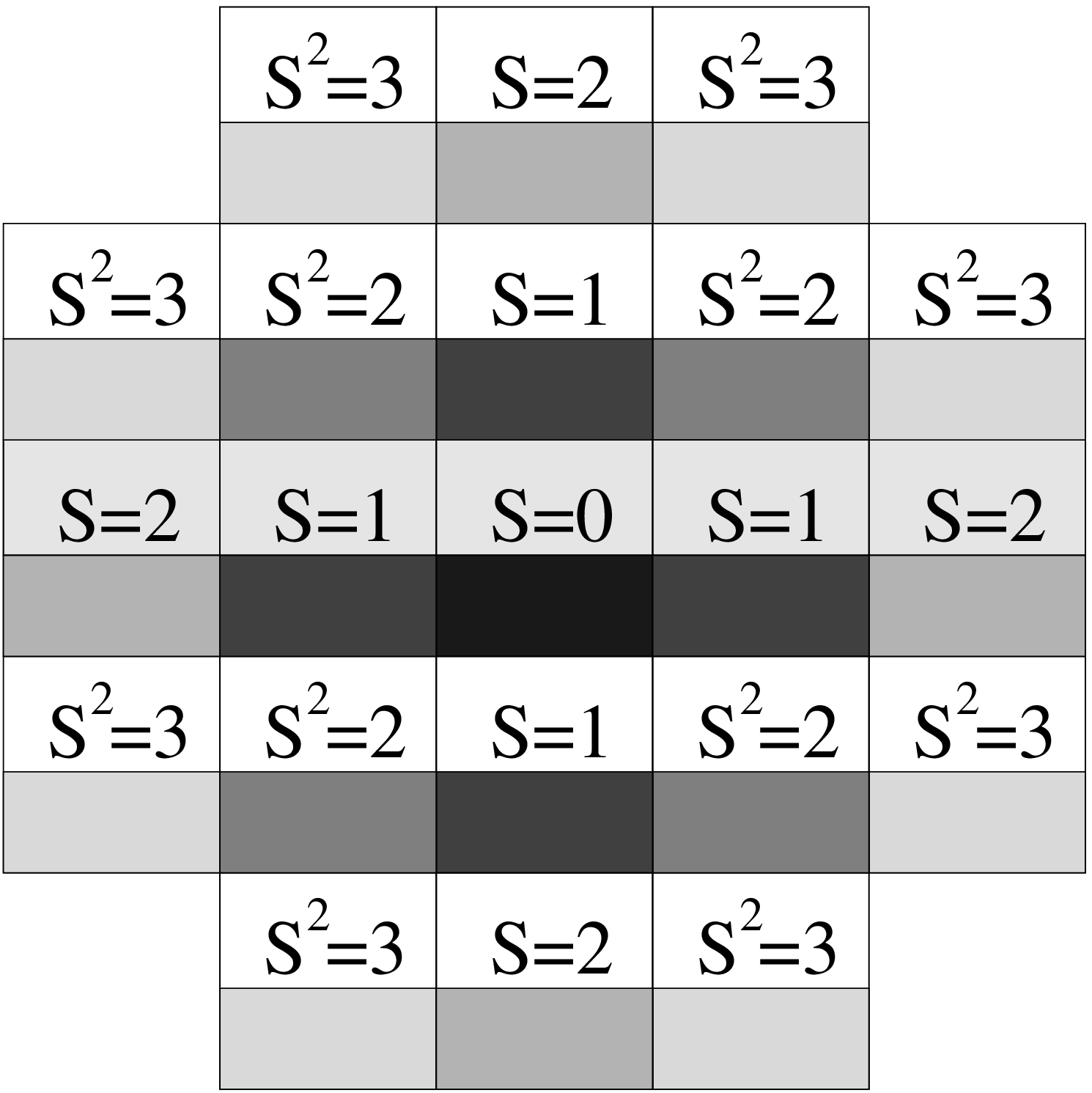}}
    \label{fig:spherical}}
  \subfigure[Schematic view of the slab--wise summation order. $n$ is
  the $z$ offset of the box, the spherical summation order in the
  $x,y$--plane is not shown.] {%
    \makebox[0.48\textwidth]{\includegraphics[width=0.3\textwidth]{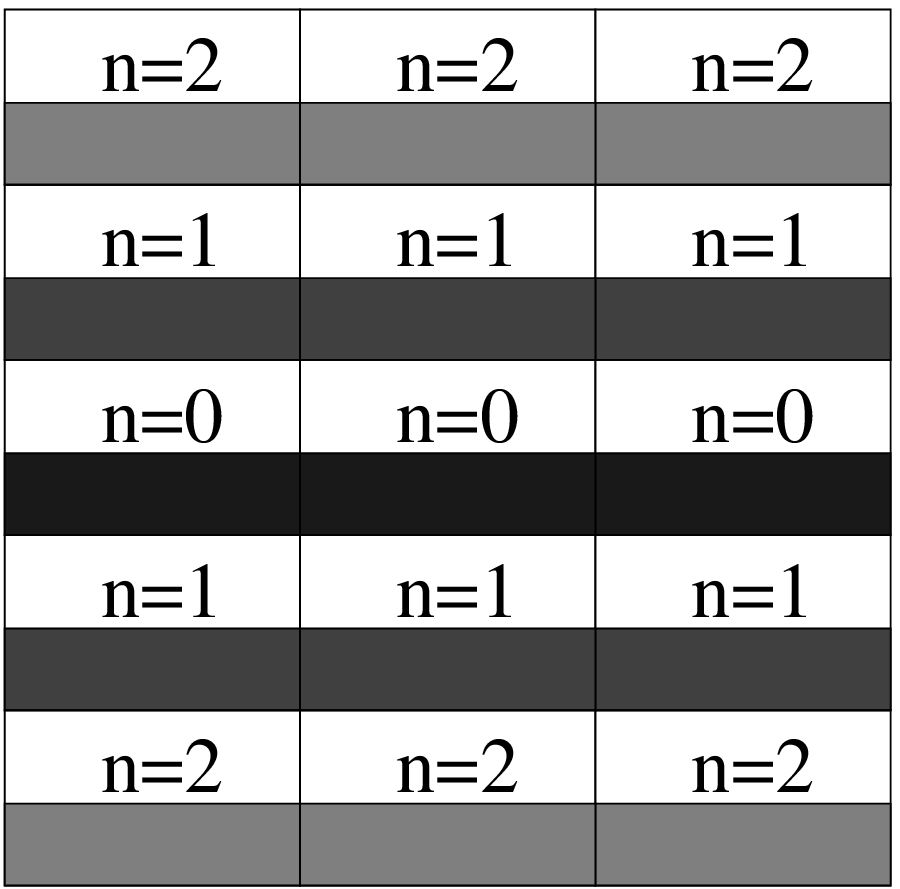}}
    \label{fig:planewise}}
\end{figure}

However, in a naive implementation, even large gap sizes will result in large
errors \cite{yeh99a}. This is due to the order of summation for the three
dimensional Coulomb sum, which is spherical by convention. This order implies
that with increasing shell cutoff $S$ the number of image shells grows faster
than the number of shells of the primary layer, namely $\O(S^3)$ versus
$\O(S^2)$ (see Fig.~\ref{fig:spherical}). In other words, we include the
unwanted terms faster than the actually wanted terms. Also the image layers
are not really infinite charged sheets but are truncated due to the cut-off.
Yeh and Berkowitz\cite{yeh99a} already suggested that this problem can be
solved by changing the order of summation. Smith has shown that by adding to
the Coulomb energy the term
\begin{equation}
  \label{eq:correction}
  E_c=2\pi M_z^2 - \frac{2\pi M^2}{3}\mcm
\end{equation}
where $M=\sum q_i p_i$ is the total dipole moment, one obtains the
result of a slab--wise summation instead of the spherical limit
\cite{smith81a}.  Slab--wise summation refers to the sum $\sum_{|n|\ge
  0} E_l(n)$, where $E_l(n)$ denotes the energy, calculated in
spherical summation order, resulting from the image layer with shift
$nL_z$ in the $z$--coordinate.  Technically this is the order where we
first treat the original layer and then add the image layers grouped
in symmetrical pairs (see Fig.~\ref{fig:planewise}).  Obviously this
summation order fits much better to the charged sheet argumentation
given above.  Although this is a major change in the summation order,
the difference given by Eq.~\eqref{eq:correction} is a very simple
term.  In fact, Smith shows that changes of the summation order always
result in a difference that depends only on the total dipole moment.

Applying this slab--wise summation order, Yeh and Berkowitz showed that a gap
size of at least $h$ is normally sufficient to obtain an moderately accurate
result. Therefore the result of a standard three dimensional summation method
plus the shape--dependent term given by Eq.~\eqref{eq:correction}, which we
refer to as a \emph{slab--wise method}, can be used to obtain a good
approximation to the result for the slab geometry with the same computational
effort as for the underlying three dimensional summation method (no matter if
a simple or sophisticated method is used). One drawback is that no theoretical
estimates exist for the error introduced by the image layers.  Therefore one
might be forced to use even larger gaps to assure that no artifacts are
produced by the image layers. One simple deducible artifact is that the
pairwise error will be position dependant.  Particles in the middle of the
slab will see no effect of the image layers due to symmetry, and particles
near the surface will encounter for the same reason the largest errors, which
is definitely an unwanted feature for studying surface effects.  Therefore
averaging error measures like the commonly used RMS force error should not be
applied without additional checks for the particles near the surfaces.

The other drawback is that normally the box now will have a significantly
larger $L_z/L$.  But at least for Ewald type methods the computation time is
proportional to this fraction.  This is easy to see as the number of
$k$--space vectors in the $z$ direction must be proportional to $L_z$ to
maintain a fixed resolution and therefore error.  It is verified
experimentally that a gap of at least $h$ is needed.  For a cubic system $h=L$
therefore the computation time at least doubles.

Nevertheless because of the bad scaling of the known methods for slab
geometries like the one by Parry \cite{parry75a,parry76a} ($\O(N^2)$) or
\algo{MMM2D} \cite{arnold01b,arnold01c} ($\O(N^{5/3})$), for particle numbers
above $N \approx 1000$ using slab--wise methods is a great improvement.

\section{The electrostatic layer correction term}
We will now derive a term that allows to calculate the \emph{exact}
contribution of the image layers very efficiently, which we will call the
electrostatic layer correction (ELC) in the following.  For the following
analysis there is no special restriction on $h$ except for $h < L_z$, which is
true even if the $L\times L\times L_z$--box is completely filled.

The method presented here is heavily based on parts of \algo{MMM2D}
\cite{arnold01b}.  We start with a formal definition of the Coulomb
energy of the slab system
\begin{equation}
  \label{eq:energy}
  E = \frac{1}{2}
  \sum_{S=0}^{\infty}
  \sum_{\substack{n \in \Z^2\times \{0\}\\n_x^2+n_y^2=S}}
  \sump_{i,j=1}^{N}
  \frac{q_{i}q_{j}}{|p_i - p_j + \Lambda n|}\mpt
\end{equation}
$\Lambda=\text{diag}(L,L,L_z)$ is a diagonal matrix describing the
shape of the box. The image boxes are denoted with the vector
$n=(n_x,n_y,n_z)$, where $n_z=0$ for now. The prime on the inner
summation indicates the omission of the self--interaction $i=j$ in the
primary box $n=(0,0,0)$ (i.\ e. the singular case). For the surrounding
dielectric medium we assume vacuum boundary conditions.

We now expand the system to a fully three-dimensional periodic system, where
$L_z$ determines the period in the $z$-coordinate as in the previous section.
We can rewrite the energy as
\begin{equation}
  \label{eq:eparts}
  E = E_s+E_c+E_{lc}\mcm
\end{equation}
where
\begin{equation}
  E_s = \frac{1}{2}\sum_{S=0}^\infty \sum_{\substack{n\in\Z^3\\n^2=S}}
  \sump_{i,j=1}^{N}
  \frac{q_{i}q_{j}}{|p_i - p_j + \Lambda n|}\mpt
\end{equation}
denotes the standard three-dimensional Coulomb--sum with spherical limit. To
evaluate this expression one can use any of the efficient algorithms, starting
with the classical Ewald summation up to modern methods like fast multipole
methods \cite{greengard87a} or mesh based algorithms\cite{deserno98a}.  $E_c$
again denotes the shape--dependent term given by Eq.~\eqref{eq:correction} and
finally
\begin{equation}
  \label{eq:elcbf}
  E_{lc} = -\frac{1}{2}\sum_{\substack{T\in\Z\\T > 0}}\sum_{n_z=\pm T}
  \sum_{S=0}^\infty\sum_{\substack{n\in\Z^2\times\{n_z\}\\n^2=S}}
  \sum_{i,j=1}^{N}
  \frac{q_{i}q_{j}}{|p_i - p_j + \Lambda n|}\mpt
\end{equation}
denotes the contribution of the image layers, for which we are going to derive
a new expression in the following.

We start with the expression for the energy induced by an image layer at
$z$--offset $n_z\neq 0$:
\begin{equation}
  \label{eq:exact}
  E_l(n_z) = -\frac{1}{2}\sum_{S=0}^\infty
  \sum_{\substack{n\in\Z^2\times\{n_z\}\\n^2=S}}
  \sum_{i,j=1}^{N}
  \frac{q_{i}q_{j}}{|p_i - p_j + \Lambda n|}\mpt
\end{equation}

It can be shown rigorously, although this is non--trivial, that
\begin{equation}
  \label{eq:convfac}
  E_l(n_z) = -\frac{1}{2}\lim_{\beta\rightarrow 0}
  \sum_{n\in\Z^2\times\{n_z\}}
  \sum_{i,j=1}^{N}
  \frac{q_{i}q_{j}e^{-\beta|p_i - p_j + \Lambda n|}}
  {|p_i - p_j + \Lambda n|}\mpt
\end{equation}
This is a convergence factor approach with a convergence factor of
$e^{-\beta|p_i - p_j + \Lambda n|}$. Note that this approach is exact only for
two-dimensional systems, for three-dimensional systems Eqs.~\eqref{eq:exact}
and \eqref{eq:convfac} differ by a multiple of the dipole moment
\cite{deleeuw80a,deleeuw80b}.

In \cite{arnold01a,arnold01b} one can find a proof for this equation and an
efficient way of calculating $E_l$ for charge neutral systems.  We do not want
to go through the full derivation again; it consists of the application of
Poisson's summation formula along both periodic coordinates and performing the
limit $\beta\rightarrow 0$ analytically. One obtains
\begin{equation}
  \label{eq:fullfar}
  E_{lc}(n_z)=-\frac{1}{2}\sum_{i,j=1}^Nq_iq_j\phi(p_i-p_j+\Lambda n)\mcm
\end{equation}
where $\phi$ is given by
\begin{multline}
  \label{eq:pfullfar}
  \phi(x,y,z) =\,
  \frac{4}{L}\sum_{\substack{k_\parallel\in\Z^2\\k_x,k_y>0}}
  \frac{e^{-2\pi |k_\parallel||z|/L}}
  {|k_\parallel|}
  \cos(2\pi k_x x/L)\cos(2\pi k_y y/L)\, +\\
  \frac{2}{L}\left(
    \sum_{k_y>0}
    \frac{e^{-2\pi k_y|z|/L}}{k_x}
    \cos(2\pi k_y y/L)\,+
    \sum_{k_x>0}
    \frac{e^{-2\pi k_x|z|/L}}{k_y}
    \cos(2\pi k_x x/L)\right)\,
  - \frac{2\pi}{L^2} |z|\mpt
\end{multline}
$k_\parallel=(k_x,k_y)$ is a Fourier variable with integer values.  $\phi$ is
an artificial pairwise potential that yields the total Coulomb energy and its
derivative produces the pairwise forces for the periodic system.

For now we only have a formula for the contribution of one image
layer, so we still have to sum over all $n_z$. This task can be
performed analytically. The terms $2\pi|z|/L^2$ can be omitted since
they are exactly the homogeneous sheet potential and we have seen
before that this cancels out for charge neutral systems (see Eq.
\eqref{eq:parpltappr}).

The summation over $n_z$ of the remaining sums over $p$ and $q$ is
fairly easy to perform using the geometric series (as these sums are
absolutely convergent, exchanging the summation over $n_z$ and the
summations over $(k_x,k_y)$ is possible). Combining the terms for
$\pm n_z$ again we obtain
\begin{equation}
  \label{eq:elc}
  E_{lc}=\sum_{i,j=1}^Nq_iq_j\psi(p_i-p_j+\Lambda n)\mcm
\end{equation}
where
\begin{equation}
  \label{eq:pelc} 
  \begin{split}
    \psi(x,y,z)=&\frac{4}{L}\sum_{k_x,k_y>0}
    \frac{\cosh(2\pi k_{\parallel }z_{ij}/L)}
    { k_{\parallel }(e^{2\pi k_{\parallel }L_{z}/L}-1)}
    \cos (2\pi k_{x}x_{ij}/L)\cos (2\pi k_{y}y_{ij}/L)+\\
    &\frac{2}{L}\sum_{k_{x}>0}\frac{\cosh(2\pi k_{x}z_{ij}/L)
      \cos(2\pi k_{x}x_{ij}/L)}{
      k_{x}(e^{2\pi k_{x}L_{z}/L}-1)}+\\
    &\frac{2}{L}\sum_{k_{y}>0}\frac{\cosh(2\pi k_{y}z_{ij}/L)\cos
      (2\pi k_{y}y_{ij}/L)}{ k_{y}(e^{2\pi k_{y}L_{z}/L}-1)}\mpt
\end{split}
\end{equation}

The forces can be obtained from that by simple differentiation since
the sums are absolutely convergent.  Although the form in
Eq.\eqref{eq:pelc} has a much better convergence than the original
form in Eq.\eqref{eq:elcbf}, its main advantage is a linear
computation time with respect to the number of particles $N$. To see
this, the equation has to be rewritten using the addition theorems for
the cosine and the hyperbolic cosine. For each $k_\parallel$ one
first calculates the sixteen terms
\begin{equation}
  \begin{split}
    \chi_{(c/s,c/s,c/s)}&=\sum_{i=1}^{N}q_{i}
    \cosh/\sinh(2\pi k_\parallel z_i/L)
    \cos/\sin(2\pi k_x x_i/L)
    \cos/\sin(2\pi k_y y_i/L) \mcm \\
    \chi_{(x,c/s,c/s)}&=\sum_{i=1}^{N}q_{i}
    \cosh/\sinh(2\pi k_x z_i/L)
    \cos/\sin(2\pi k_ x x_i/L) \mcm \\
    \chi_{(y,c/s,c/s)}&=\sum_{i=1}^{N}q_{i}
    \cosh/\sinh(2\pi k_y z_i/L)
    \cos/\sin(2\pi k_y y_i/L)\mcm
  \end{split}
\end{equation}
where the indices in the obvious way determine which of the functions
cosine (hyperbolicus) or sinus (hyperbolicus) are used.
Then we evaluate
\begin{equation}
  \begin{split}
    E_{lc}=&\frac{4}{L}\sum_{k_x,k_y>0}
    \frac{1}{(e^{2\pi k_\parallel L_z/L}-1)k_\parallel}\Bigl(
    \chi_{(ccc)}^2 + \chi_{(csc)}^2 + \chi_{(ccs)}^2 + \chi_{(css)}^2 - \\
    &\phantom{\frac{4}{L}\sum_{k_x,k_y>0}
      \frac{1}{(e^{2\pi k_\parallel L_z/L}-1)k_\parallel}\Bigl(}
    \chi_{(scc)}^2 - \chi_{(ssc)}^2 - \chi_{(scs)}^2 - \chi_{(sss)}^2\Bigr) +\\
    &\frac{2}{L}\sum_{k_x > 0}
    \frac{1}{(e^{2\pi k_x L_z/L}-1)k_x}\Bigl(
    \chi_{(xcc)}^2 + \chi_{(xcs)}^2 - \chi_{(xsc)}^2 - \chi_{(xss)}^2\Bigr) + \\
    &\frac{2}{L}\sum_{k_y > 0}
    \frac{1}{(e^{2\pi k_y L_z/L}-1)k_y}\Bigl(
    \chi_{(ycc)}^2 + \chi_{(ycs)}^2 - \chi_{(ysc)}^2 - \chi_{(yss)}^2\Bigr)
  \end{split}
\end{equation}

Similar expansions using the same sixteen terms can also be found for
the forces.  Obviously this has linear computation time with respect
to the number of particles, as the only summations over all the
particles occur in the $\chi_*$.  But up to now there is still the infinite
summation over $k_\parallel$. So the next task is to derive an
estimate for the error induced by the replacement of the infinite sum
by a finite one.

\section{Error estimates}

Since $E_{lc}$ is written as sum over an alternative potential $\psi$, it is
reasonable to derive an upper bound for the error from the calculation of
$\psi$ only with a finite cutoff. From this upper bound, crude estimates for
other error measures such as the RMS (root-mean-square) force error can be
derived. Again these error estimates are taken from \algo{MMM2D}
\cite{arnold01a,arnold01b}. As we will show later, the error distribution is
not uniform along the $z$--axis. The error gets maximal for particles near the
borders. Since these particles will normally be those of special interest, the
maximal pairwise error should be some magnitudes smaller than the thermal
noise.

While the error bounds for \algo{MMM2D} were only used to tune the algorithm,
the error estimates for $E_{lc}$ can also be used to obtain an error bound for
the slab--wise method from Ref. \cite{yeh99a}, and hence one can determine ``a
priori'' the necessary gap size to reach a preset precision.  Therefore we
also have to deal with small cutoffs, especially the case when no terms of
$E_{lc}$ are added.

First we choose a cutoff $R\ge 1$ and then evaluate $E_{lc}$ only over
the area
\begin{equation}
  \label{defgammar}
  \begin{split}
    \Gamma_R=&\left\{(k_x,k_y)\in \Z^2\,|\,
      k_x,k_y > 0, (k_x-1)^2 + (k_y-1)^2 < R^2\right\}\cup\\
    &\left\{(k_x,0)\in\Z\times\{0\}\,|\, k_x < R\right\}\\
    &\left\{(0, k_y)\in\{0\}\times\Z\,|\, k_y < R\right\}\mpt
  \end{split}
\end{equation}
The three sets correspond to the three sums in Eq.\eqref{eq:pelc}. Therefore
we actually evaluate
\begin{equation}
  \begin{split}
    E_{lc}=&\frac{4}{L}
    \sum_{\substack{k_x,k_y>0,\\(k_x-1)^2 + (k_y-1)^2 < R^2\\}}
    \frac{\cosh(2\pi k_{\parallel }z_{ij}/L)}
    {k_{\parallel }(e^{2\pi k_{\parallel }L_{z}/L}-1)}
    \cos (2\pi k_x x_{ij}/L)\cos (2\pi k_y y_{ij}/L)+\\
    &\frac{2}{L}\sum_{0 < k_x < R}
    \frac{\cosh(2\pi k_xz_{ij}/L)\cos(2\pi k_x x_{ij}/L)}
    {k_{x}(e^{2\pi k_x L_{z}/L}-1)}+\\
    &\frac{2}{L}\sum_{0< k_y < R}
    \frac{\cosh(2\pi k_y z_{ij}/L)\cos(2\pi k_ y y_{ij}/L)}
    {k_{y}(e^{2\pi k_y L_{z}/L}-1)}\mpt
  \end{split}
\end{equation}

$\Gamma_R$ may look more complicated then necessary.  But this form
enables us to find a rigorous upper bound for the error.  An upper
bound for the absolute value of the summands is
\begin{multline}
  \label{eq:sumapprox}
    \left|\frac{\cosh(2\pi k_{\parallel }z_{ij}/L)}
      {k_{\parallel }(e^{2\pi k_{\parallel }L_{z}/L}-1)}
      \cos (2\pi k_x x_{ij}/L)\cos (2\pi k_y y_{ij}/L)\right|\le \\
    e^{-2\pi k_\parallel Lz/L}\frac{\cosh(2\pi k_\parallel z_{ij}/L)}
    {(1 - e^{-2\pi k_\parallel L_z/L})k_\parallel}\le
    e^{-2\pi k_\parallel Lz/L}\frac{\cosh(2\pi k_\parallel h/L)}
    {(1 - e^{-2\pi k_\parallel L_z/L})k_\parallel}\mpt
\end{multline}
Of course because the cosine hyperbolicus is monotonous, one could
use any larger value for $h$. This is for example necessary in a priori
estimations.  Using this we find the upper bound for the maximal
pairwise error in the potential by a simple approximation of the sums
by integrals as
\begin{equation*}
  \label{eq:sumfull}
  \tau_E:=\frac{1/2+ (\pi R)^{-1}}{e^{2\pi R L_z/L} - 1}
  \left(\frac{\exp(2\pi R h/L)}{Lz - h} +
    \frac{\exp(-2\pi R h/L)}{Lz + h}\right)\mpt
\end{equation*}
Details can again be found in \cite{arnold01a}. By an
analogous derivation one can also find an upper bound on the maximal
pairwise error on each of the three force components as
\begin{equation}
  \begin{split}
    \tau_F:=\frac{1}{2(e^{2\pi R L_z/L} - 1)}\Biggl(
      &\left(\frac{2\pi R + 4}{L} + \frac{1}{L_z - h}\right)
      \frac{\exp(2\pi R h/L)}{(L_z - h)} + \\
      &\left(\frac{2\pi R + 4}{L} + \frac{1}{L_z + h}\right)
      \frac{\exp(-2\pi R h/L)}{(L_z + h)}\Biggr)\mpt
  \end{split}
\end{equation}
This is also an weaker bound for the potential. In other words, the
maximal pairwise error on the forces is larger than the error in the
potential. For $R=1$ one obtains an overall estimate of the magnitude
of the contribution of the image layers, i.\ e. an error estimate for
the slab--wise methods.

Note that Eq.\eqref{eq:sumapprox} shows that the error in the
potential or the force for a single particle will be largest if it is
located near the gap, since there $|z_{ij}|$ will be maximal.  This
effect will increase with increasing $R$.  Therefore when using the
layer correction one must apply non--averaging error estimates such as
our maximal pairwise error.  Averaging error estimates such as the RMS
force error might be misleading about the error on the particles of
interest.

Moreover the error will decrease exponentially with the distance from
the gap.  Since the particles near the gap (i.e. the surface) are
normally of special interest in simulations with slab geometry,
averaging error measures like the RMS force error might be misleading
about the effect of these errors.

All our error estimates show that the error drops exponential both with $R$
and $L_z/L$. The decay in $R$ means that it is easy to achieve high accuracies
with our layer correction formula, while the decay in $L_z/L$ shows that
slab--wise methods can achieve good accuracies without increasing $L_z/L$
too much.

Although we do not encourage using the RMS error measure as we
explained above, we still want to give an upper bound on the average
error in $E_{lc}$ and the RMS force error.  We assume that the
pairwise potentials of the different particle pairs are independent
identically distributed random variables, which is true for homogenous
random systems and normally a good assumption otherwise.  Of course
the self interaction of the particles, i.\ e.  $q_i^2\psi(0,0,0)$ has
to be omitted.  Let $\sigma_E$ be the variance of this random
variable, then it is easy to see that $\sigma_E\le\tau_E^2$.  Using
this one can show that
\begin{equation}
  \label{eq:avgelc}
  \left<E_{lc} - \sum_{i}q_i^2\psi(0,0,0)\right> \le
  Q^2\sqrt{\sigma_E}\le Q^2\tau_E\mcm
\end{equation}
where $Q:=\sum_{i}q_i$.

Similarly we define $\sigma_F$ as the variance of the forces measured
in the Euclidian norm. Then because of the component--wise maximal error
estimate for the force, we have $\sigma_F^2\le 3\tau_F^2$, and one obtains
\begin{equation}
  \label{eq:avgrms}
  \left<\sqrt{1/N\sum_{i}|\Delta F_{lc}^i|}\right> \le
  \sqrt{3}Q^2/\sqrt{N}\tau_F\mcm
\end{equation}
where $\Delta F_{lc}^i$ denotes the error in the layer correction
force on particle $i$. Note that both estimates \eqref{eq:avgelc} and
\eqref{eq:avgrms} are much larger than the real error as one
expects $\sigma$ to be \emph{much} smaller the maximal error (about
$2-4$ magnitudes).

\section{Numerical demonstration}

\begin{figure}[htbp]
  \begin{center}
    \includegraphics[width=\picturewidth]{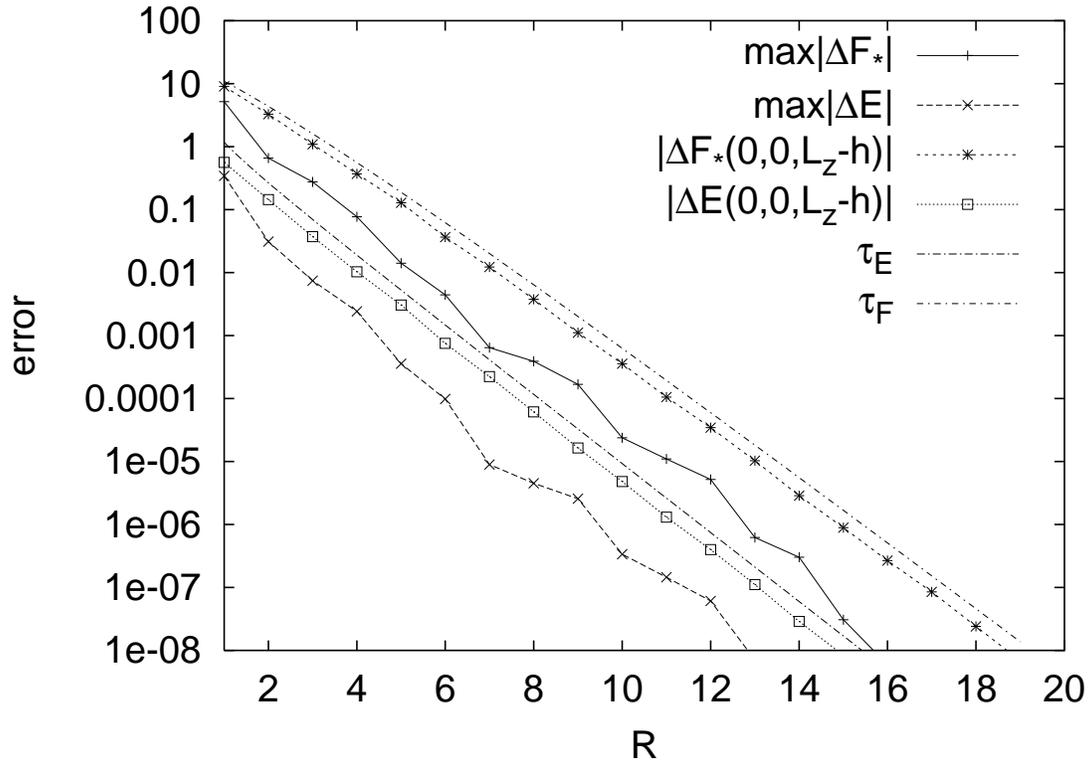}
    \caption{The measured error in the potential and the force of the
      \algo{ELC}--term versus the estimates $\tau_E$, $\tau_F$ for different cutoff radii $R$}
    \label{fig:pw}
  \end{center}
\end{figure}

In this section we want to show results from our implementation of the
layer correction term (\algo{ELC}). First we want to show that our
maximal pairwise error bounds are correct. To this aim we place two
particles randomly in a box of size $1\times 1\times 0.8$, so that we
leave a gap of $\delta=0.2$ in a box of dimensions $1\times 1\times
1$.  Fig.~\ref{fig:pw} shows the maximal potential and force error
that occurred during 10000 evaluations and our estimates $\tau_E$ and
$\tau_F$.  Moreover we have included the result for a particle pair
with a relative position of $(0,0,h)$, the worst case position.  For
such a position the error estimate is exact up to the approximation of
the sum by an integral.  As ``exact'' force we used $R=30$.

One can see that the maximal error estimates are always above the
measured deviations.  Even after 10000 random evaluations the maximum
error is considerably lower than for the special pair with relative
position $(0,0,h)$, and the error is still not very smooth.  This
effects are due to the Fourier representation with exponentially
decaying coefficients, which makes the worst case error extremely rare.
But in a real simulation the particle distribution is not necessarily
homogeneous and to be on the safe side one has to deal also with the
rare worst case error. Nevertheless Fig.~\ref{fig:pw} shows that the
error coming from the image layers can be strictly controlled.

\begin{figure}[htbp]
  \subfigure[Computation times for the ELC term for different system sizes]{%
    \makebox[\picturewidth]{\includegraphics[width=\picturewidth]{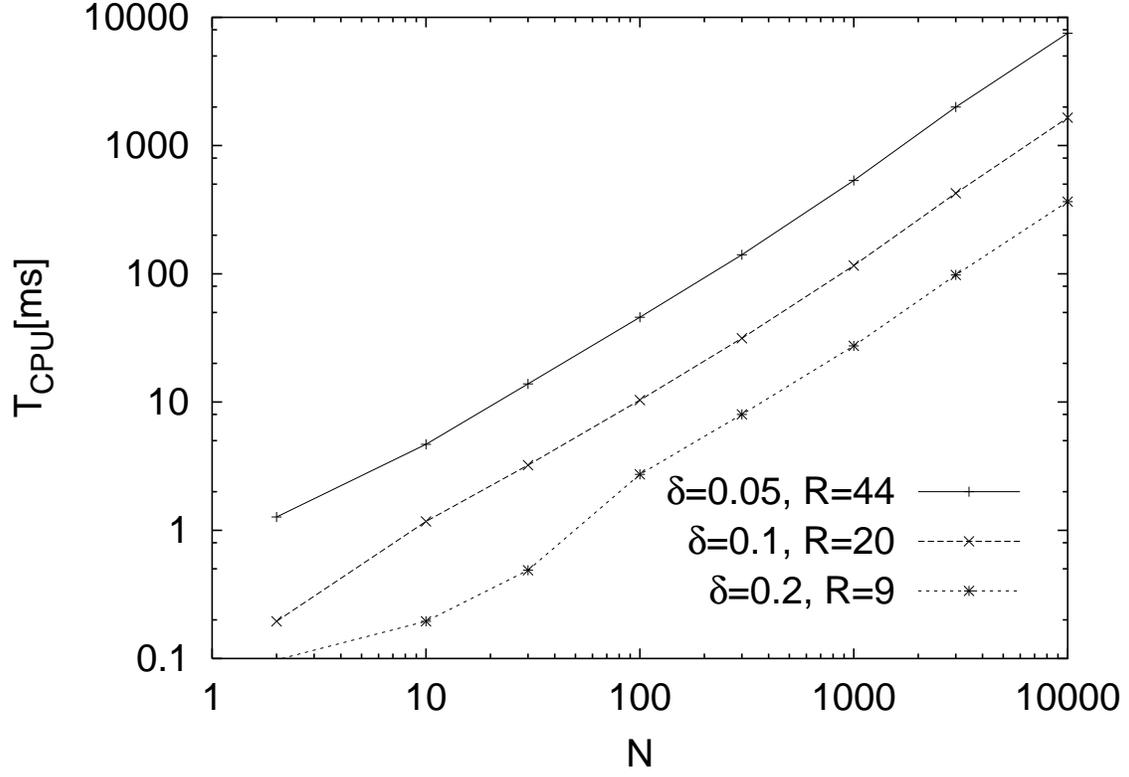}}
    \label{fig:cpu}}
  \subfigure[RMS force errors for these systems]{%
    \makebox[\picturewidth]{\includegraphics[width=\picturewidth]{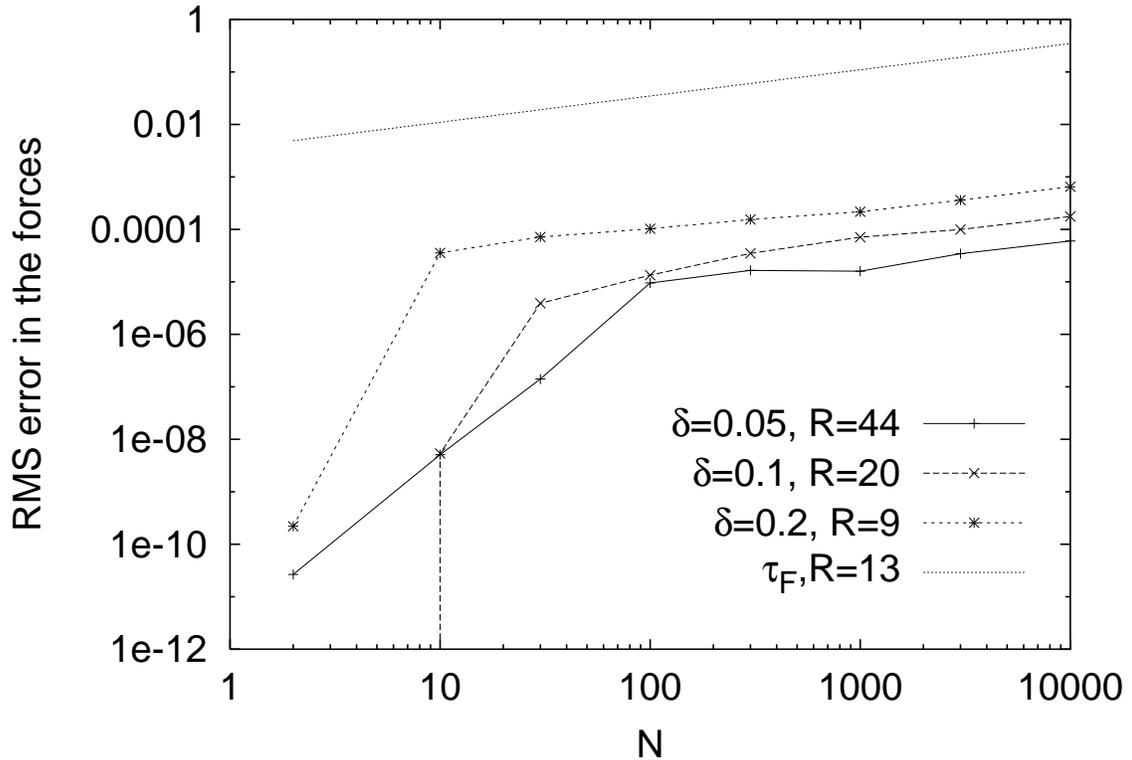}}
    \label{fig:err}}
  \caption{Results for homogeneous random systems of $N$ particles
    for different gap sizes $\delta$}
\end{figure}

Next we investigate the computation times of our implementation of the
ELC term.  For different gap sizes $\delta=0.05,0.1$ and $0.2$ we show
in Fig.~\ref{fig:cpu} the computation time $T_{CPU}$ for different
numbers of particles $N$. The systems were again consisting of
uniformly randomly distributed particles of charges $\pm 1$. The
maximal pairwise error was fixed to be $10^{-4}$. The times are
averages over 10 runs on a Compaq XP1000.  The computation time for
the same system consisting of 1000 charges using P$^3$M is $\approx
330ms$ for a typical RMS force error of $10^{-4}$.  Therefore even the
small gap of $0.05$ gives just the same order of computation time. For
the normally used gap sizes of $\approx 0.2 L_z$ the computation time
is negligible compared to P$^3$M.  In Fig.~\ref{fig:err} we show the RMS
force errors that occurred.  One can see the predicted $Q^2/\sqrt{N}$
behavior. We also show the theoretical upper bound
$\sqrt{3}Q^2/\sqrt{N}\tau_F$ for $R=13$ and $\delta=0.2$, which is
considerably above as expected.

\begin{figure}[htbp]
  \begin{center}
    \includegraphics[width=\picturewidth]{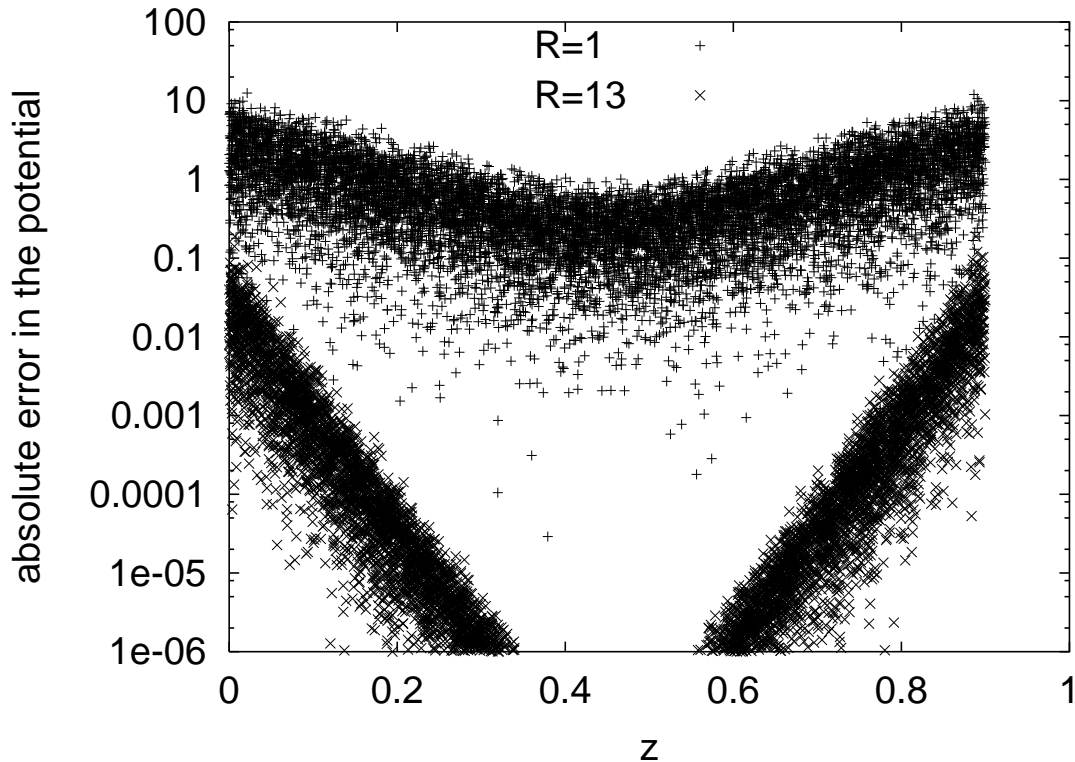}
    \caption{Error distribution of the layer correction along the
      $z$--axis for 100 random systems with 100 particles}
    \label{fig:zerr}
  \end{center}
\end{figure}

As the last important fact we demonstrate that the errors for the
layer correction indeed are maximal near the gap (i.e. near the
surface). For Fig.~\ref{fig:zerr} 100 particles were put 100 times
randomly in a box with a gap of $\delta=0.1$. Then for every particle
the magnitude of the layer correction for $R=40$, which is a good
approximation to the full $E_{lc}$, and the difference in the layer
correction between $R=5$ and $R=40$ was drawn against the
$z$-coordinate.  Clearly the error always is largest near the gap.
This effect increases with increasing $R$, which is easy to understand
from the error formula. Therefore the full RMS error of the system
might be completely misleading about the effect the errors have on the
particles near the gap as we mentioned before.  Nevertheless the
figure shows that $E_{lc}$ with $R=5$ reduces the error near the
surfaces by a factor of $\approx 100$.

\section{Conclusion}

We have derived a term called \algo{ELC} to efficiently calculate the
contribution of the image layers in three dimensionally periodically
replicated slab systems.  \algo{ELC} scales as the number $N$ of particles and
has a rigorous error bound.  Moreover this error bound can be used to estimate
the size of the image layer contribution and therefore gives a bound on the
error introduced by slab--wise methods as proposed by Yeh and Berkowitz.
We have found that the error for these methods decays exponentially in
$L_z/L$. However, the errors are not uniformly distributed over the slab,
namely they are worst at the surfaces of the slabs. This strongly suggests to
restrict the maximal pairwise error instead of the usually assumed RMS-errors.

In a forthcoming paper \cite{dejoannis02a} we will focus on the
application of \algo{ELC} to the standard Ewald method and to
\algo{P$^3$M}.  We will show how the error formulas of Kolafa and
Perram \cite{kolafa92a} have to be adapted to allow non--cubic
simulation boxes which is essential for using \algo{ELC} with $R=1$,
i.\ e. a slab--wise method.  For all combinations we present numerical
results which allow easily to decide which method is optimal for use
in a real simulation.

\section*{Acknowledgments}
Financial support from the DFG ``Schwerpunkt Polyelektrolyte'' is gratefully
acknowledged. 
\bibliographystyle{aip}


\end{document}